 \documentclass[final,5p,times,twocolumn,authoryear]{elsarticle}

\biboptions{square,comma,sort&compress}
\setcitestyle{numbers}

\usepackage{amssymb}
\usepackage{lipsum}
\usepackage{graphicx}
\usepackage{dcolumn}
\usepackage{bm}

\usepackage{bm, color}
\usepackage{lipsum}
\usepackage[caption=false]{subfig}
\usepackage{dblfloatfix}
\usepackage{orcidlink}

\definecolor{ryangreen}{rgb}{0.00,0.0,0.0}

\definecolor{red}{rgb}{0.81,0.13,0.16}

\usepackage{hyperref}
\hypersetup{
    colorlinks=true,
    linkcolor=blue,
    filecolor=magenta,      
    urlcolor=blue,
    citecolor=blue,
    }
\usepackage[all]{hypcap}
\usepackage{amsmath}

\journal{Physics Letters B}

\begin{document}

\begin{frontmatter}

\title{Quantum Monte Carlo Calculations of Light Nuclei\\ with Fully Propagated Theoretical Uncertainties}


\author[label1,label2]{Ryan Curry\orcidlink{0000-0002-8013-2374}}
\ead{rmcurry@lanl.gov}
\author[label3,label4,label5]{Kai Hebeler\orcidlink{0000-0003-0640-1801}}
\author[label2]{Stefano Gandolfi\orcidlink{0000-0002-0430-9035}}
\author[label1]{Alexandros Gezerlis\orcidlink{0000-0003-2232-2484}} 
\author[label3,label4,label5]{\\ Achim Schwenk\orcidlink{0000-0001-8027-4076}}
\author[label2]{Rahul Somasundaram\orcidlink{0000-0003-0427-3893}}
\author[label2]{Ingo Tews\orcidlink{0000-0003-2656-6355}}
\affiliation[label1]{organization={Department of Physics, University of Guelph, Guelph, Ontario N1G 2W1, Canada}}
\affiliation[label2]{organization={Theoretical Division, Los Alamos National Laboratory, Los Alamos, New Mexico 87545, USA}}
\affiliation[label3]{organization={Technische Universit\"at Darmstadt, Department of Physics, 64291 Darmstadt, Germany}}
\affiliation[label4]{organization={ExtreMe Matter Institute EMMI, GSI Helmholtzzentrum f\"ur Schwerionenforschung GmbH, 64291 Darmstadt, Germany}}
\affiliation[label5]{organization={Max-Planck-Institut f\"ur Kernphysik, Saupfercheckweg 1, 69117 Heidelberg, Germany}}


\begin{abstract}
We report on the first quantum Monte Carlo calculations of helium isotopes with fully propagated theoretical uncertainties from the interaction to the many-body observables. 
To achieve this, we build emulators for solutions to the Faddeev equations for the binding energy and Gamow-Teller matrix element of $^3\text{H}$, as well as for auxiliary-field diffusion Monte Carlo calculations of the $^4\text{He}$ charge radius, employing local two- and three-body interactions up to next-to-next-to-leading order in chiral effective field theory. 
We use these emulators to determine the posterior distributions for all low-energy couplings that appear in the interaction up to this order using Bayesian inference while accounting for theoretical uncertainties.
We then build emulators for auxiliary-field diffusion Monte Carlo for helium isotopes and propagate the full posterior distributions to these systems.
Our approach serves as a framework for $\textit{ab initio}$ studies of atomic nuclei with consistently treated and correlated theoretical uncertainties.
\end{abstract}







\end{frontmatter}




\section{Introduction }\label{Xsec1-1}
To study the nuclear many-body problem, one needs to employ both a computational approach for solving the many-body Schr\"{o}dinger equation and an accurate description of the interactions between nucleons.
Currently, there are a number of \textit{ab initio} many-body approaches that reliably describe different nuclear systems
\cite{Hagen_Ekstrom_Forssen_etal_2016,Barrett_Navratil_Vary_2013,Carbone_Cipollone_Barbieri_etal_2013,Hergert_Bogner_Morris_etal_2016,Lynn_Tews_Gandolfi_etal_2019,Drischler_Holt_Wellenhofer_2021,Lee_2025}. Typically these employ interactions derived from chiral effective field theory (EFT) \cite{Epelbaum_Hammer_Meissner_2009, Machleidt_Entem_2011},
which is a systematically improvable theory for nuclear forces.
And while there remain open questions  \cite{Nogga_Timmermans_vanKolck_2005, Tews_Davoudi_Ekstrom_etal_2020}, many-body calculations employing chiral EFT interactions have been very successful \cite{Navratil_Gueorguiev_Vary_etal_2007,Hagen_Papenbrock_Dean_etal_2008,Epelbaum_Krebs_Lee_etal_2011,Hergert_Binder_Calci_etal_2013,Gezerlis_Tews_Epelbaum_etal_2013,Pastore_Pieper_Schiavilla_etal_2013,Bacca_Barnea_Hagen_etal_2013,Soma_Cipollone_Barbieri_etal_2014,Ekstrom_Jansen_Wendt_etal_2015,Tews_Carlson_Gandolfi_etal_2018,Morris_Simonis_Stroberg_etal_2018,Drischler_Hebeler_Schwenk_2019,Stroberg_Holt_Schwenk_etal_2021,Hu_Jiang_Miyagi_etal_2022,King_Baroni_Cirigliano_etal_2023,Miyagi_Cao_Seutin_etal_2024,Marino_Jiang_Novario_2024,Heinz_Miyagi_Stroberg_etal_2025,Arthuis_Hebeler_Schwenk_2024,Bonaiti_Hagen_Papenbrock_2025,Li_Miyagi_Schwenk_2025,Kuske_Miyagi_Arcones_etal_2025}.

One of the current challenges in nuclear theory is how to properly treat the theoretical uncertainties that stem from our incomplete description of nuclear interactions \cite{Epelbaum_Krebs_Meissner_2015,Melendez_Furnstahl_Phillips_etal_2019,Epelbaum_Golak_Hebeler_etal_2019,Drischler_Melendez_Furnstahl_etal_2020,Hu_Jiang_Miyagi_etal_2022,Armstrong_Giuliani_Godbey_etal_2025,Plies_Heinz_Schwenk_2025}.
The standard approach is to perform many-body calculations of the observable of interest at successive orders in the chiral expansion and use these results to estimate the effects of the EFT truncation.
In this Letter, we adopt a different approach to estimating these uncertainties in a many-body calculation: we first estimate the truncation uncertainty of the EFT expansion at the level of the interaction, and then propagate these to the many-body system.
To do this, one requires the ability to repeat calculations a large number of times, which, at present, is unfeasible for few- and many-body systems.
One solution to this problem is the use of emulators \cite{Frame_He_Ipsen_etal_2018, Konig_Ekstrom_Hebeler_etal_2020, Bonilla_Giuliani_Godbey_etal_2022,Duguet_Ekstrom_Furnstahl_etal_2024,Cook_Jammooa_Hjorth-Jensen_etal_2025}.
Here, we develop a framework for the  propagation of truncation uncertainties at next-to-next-to-leading order (N$^2$LO), and study observables and their correlations for the helium isotopes.

Emulators allow for a rapid evaluation of nuclear observables for any set of values of the low-energy couplings (LECs) of a chiral EFT Hamiltonian.
This, in turn, enables us to perform a Bayesian fit of local two- and three-nucleon interactions allowing a determination of the full posterior distribution for all LECs up to N$^2$LO.
We implement both eigenvector continuation (EC) \cite{Frame_He_Ipsen_etal_2018,Duguet_Ekstrom_Furnstahl_etal_2024} and the parametric matrix model (PMM) \cite{Cook_Jammooa_Hjorth-Jensen_etal_2025} to emulate solutions to the Faddeev equations for the $^3$H ground-state energy, as well as its Gamow-Teller matrix element \cite{Gazit_Quaglioni_Navratil_2009, Gazit_Quaglioni_Navratil_2019}.
In addition, we also employ the auxiliary-field diffusion Monte Carlo (AFDMC) method \cite{Schmidt_Fantoni_1999} to fit our interaction to the charge radius of $^4$He \cite{Krauth_Schuhmann_Ahmed_etal_2021} using a PMM emulator.

\begin{figure}

\centerline{\includegraphics[width=\columnwidth]{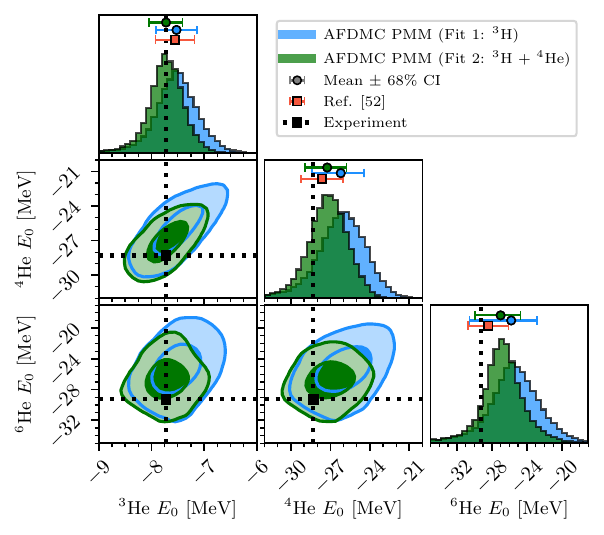}}
\caption{Corner plot for the posterior distributions of the ground-state energies of helium isotopes with fully propagated uncertainties.
    Fit 1 adjusts 3N forces to the $^3$H energy and GT matrix element, while fit 2 also includes the $^4$He charge radius.
    The 2D panels show the posterior at 50\% and 90\% credible intervals.
    We show the mean values with 68\% credible intervals for our posteriors (error bars) and compare with experiment~\cite{Wang_Huang_Kondev_etal_2021} and AFDMC calculations of Ref.~\cite{Lonardoni_Gandolfi_Lynn_etal_2018}.
    }
    \label{fig:He}
\end{figure}

For all calculations, we use a local N$^2$LO interaction at a cutoff of $R_0 = 0.6\ \text{fm}$ for nucleon-nucleon (NN) and three-nucleon (3N) contributions, with the $\text{E}_{\tau}$ form of the short-range 3N force \cite{Lynn_Tews_Carlson_etal_2016,Somasundaram_Lynn_Huth_etal_2024}.
We then build PMM emulators for AFDMC calculations of the
ground-state energies of $^3\text{He}$, $^4\text{He}$, and $^6\text{He}$ to propagate the uncertainties contained in the LEC posteriors to the final many-body predictions.
Our main results are shown in Fig.~\ref{fig:He}.
Our framework, i.e., propagating LEC uncertainties to many-body observables by using emulators, represents a systematic approach to {\it ab initio} calculations and uncertainty quantification that naturally allows for the study of correlations across different nuclear systems, and will also prove useful in future work explicitly accounting for such correlations in fits to data.

\section{Chiral EFT and three-nucleon interactions}\label{Xsec2-2}
Chiral EFT offers a systematically improvable approach to formulate nuclear interactions, casting them as an expansion in powers of $Q/\Lambda_b$, where $Q$ is the typical nucleon momentum and $\Lambda_b$ is the breakdown scale of the theory.
In practice, the expansion must be truncated at a finite order which introduces an inherent theoretical uncertainty.
By including higher-order terms in the expansion, one improves the description of nuclear interactions and reduces the theoretical uncertainty at the cost of more complicated structures and additional LECs.
Much work has been done to address the validity of estimations of truncation uncertainties~\cite{Furnstahl_Klco_Phillips_etal_2015, Melendez_Furnstahl_Phillips_etal_2019,Epelbaum_Golak_Hebeler_etal_2019}.

Chiral EFT interactions contain terms describing the long- and intermediate-range pion-exchange contributions as well as short-range contact interactions.
The contact interactions in the NN sector at N$^2$LO are described by 9 LECs
that were fit to NN phase shifts~\cite{Stoks_Klomp_Rentmeester_etal_1993} for the interactions used here~\cite{Somasundaram_Lynn_Huth_etal_2024}.
At N$^2$LO, the leading 3N interactions appear.
These include a long-range two-pion-exchange (TPE) term, an intermediate-range one-pion-exchange-contact term, as well as a pure contact term \cite{vanKolck_1994, Epelbaum_Nogga_Glockle_etal_2002}.
The TPE couplings already appear in the NN interaction, and can be determined from an analysis of pion-nucleon scattering \cite{Hoferichter_RuizdeElvira_Kubis_etal_2016}.
The values we use for these LECs at N$^2$LO are reported in Ref.~\cite{Somasundaram_Lynn_Huth_etal_2024}.
At present, we have not included the uncertainty on these couplings in our uncertainty propagation, and will explore this in future work.
The two shorter-range terms come with additional couplings, $c_D$ and $c_E$.
Typically, these 3N LECs have been fit to few-body observables~\cite{Nogga_Kamada_Glockle_etal_2002, Gazit_Quaglioni_Navratil_2009,
Hebeler_Bogner_Furnstahl_etal_2011,
Baroni_Girlanda_Kievsky_etal_2016, Lynn_Tews_Carlson_etal_2016, Epelbaum_Golak_Hebeler_etal_2019, Kravvaris_Quinlan_Quaglioni_etal_2020, Wesolowski_Svensson_Ekstrom_etal_2021}; however, they have also been fit to the properties of larger nuclei such as $^{16}\text{O}$ \cite{Ekstrom_Jansen_Wendt_etal_2015,Huther_Vobig_Hebeler_etal_2020,Hu_Jiang_Miyagi_etal_2022,Arthuis_Hebeler_Schwenk_2024}.

In this work, we fit $c_D$ and $c_E$ to the energy and Gamow-Teller (GT) matrix element of $^3\text{H}$, and the charge radius of $^4$He \cite{Gazit_Quaglioni_Navratil_2009, Wesolowski_Svensson_Ekstrom_etal_2021, Tews_Somasundaram_Lonardoni_etal_2025}.
To calculate the triton observables, we solve the Faddeev equations, which are a set of three coupled equations coming from a Lippmann-Schwinger-type approach to the three-body problem \cite{Glockle_1983, Stadler_Glockle_Sauer_1991, Glockle_Witala_Huber_etal_1996, Hueber_Glockle_Kamada_etal_1997}.
An iterative solution of the Faddeev equations leads to the exact solution of the three-body wave function, which can be used to evaluate expectation values.
We include NN partial-wave contributions up to total angular momentum $J_{\text{max}}=8$, which introduces at most a partial-wave-basis truncation error of a few keV.  As a result, we neglect this contribution in our uncertainty propagation.
To compute the GT matrix element, we use the one- and two-body currents as in Ref.~\cite{Klos_Carbone_Hebeler_etal_2017, Tews_Somasundaram_Lonardoni_etal_2025}.
We impose the same local regulator for the two-body currents as is used for the NN interaction \cite{Somasundaram_Lynn_Huth_etal_2024}.
The $^4$He charge radius $r_\text{ch}$ is computed using AFDMC by first computing the point-proton radius and then accounting for the relevant electromagnetic corrections \cite{Lonardoni_Gandolfi_Lynn_etal_2018}.
In this way, we solve for the $^3$H energy, GT matrix element, and the $^4$He $r_\text{ch}$ for a given chiral EFT interaction characterized by a specific set of 11 LECs.

To determine the posterior distributions for our LECs, we employ a Bayesian fit, defining the likelihood function,
\begin{align} \label{likelihood}
	\mathcal{L} \propto \prod_i \text{exp} \biggl[ -\frac{1}{2} \left( \frac{X_i^{\text{exp}} - X_i^{\text{theo}}}{\sigma_i} \right)^2 \biggl],
\end{align}
where the $X$ represent the observables we fit against ($^3$H energy, $^3$H GT matrix element, $^4$He $r_\text{ch}$).
The uncertainty $\sigma_i^2 = \sigma_{i,\text{exp}}^2 +  \sigma_{i,\text{theo}}^2 + \sigma_{i,\text{emulate}}^2$ includes experimental uncertainties, the average emulator uncertainties (like those shown in Fig.~\ref{fig:ntrain}), and incorporates also the chiral EFT truncation errors.
The latter are estimated using the ``EKM'' prescription~\cite{Epelbaum_Krebs_Meissner_2015, Somasundaram_Lynn_Huth_etal_2024} (with $\Lambda_b = 600$\,MeV), which requires calculations at all orders up to N$^2$LO in the chiral expansion; see Ref.~\cite{Somasundaram_Lynn_Huth_etal_2024} for more details. 
For the LO and NLO calculations of the GT matrix element, this involves using only the one-body currents from Refs.~\cite{Klos_Carbone_Hebeler_etal_2017, Tews_Somasundaram_Lonardoni_etal_2025}.
The experimental and emulator uncertainties are significantly smaller than the theoretical truncation uncertainties, and thus should have very little impact on the final posterior distributions \cite{Somasundaram_Lynn_Huth_etal_2024}. To verify this, we repeated our fits following the prescription discussed in Ref.~\cite{Somasundaram_Lynn_Huth_etal_2024} and indeed found the experimental and emulator uncertainties contribute negligibly to our final posterior distributions. This justifies our modeling $\sigma_{i}^2$ as a sum of terms in quadrature.
The prior for the 9 NN LECs is given by the posterior distribution of the NN Bayesian fit of Ref.~\cite{Somasundaram_Lynn_Huth_etal_2024}, while for $c_D$ and $c_E$ we have chosen uniform priors.

The likelihood chosen in Eq.~(\ref{likelihood}) is one possible choice for use in a Bayesian inference fit of the LECs. Formally, propagating uncertainties implies constructing the posterior predictive distribution (PPD) for an observable  $O_A$ (such as the ground-state energy) as
\begin{align}
p(O_A|D)=\int d\boldsymbol{\theta} \, p(O_A|\boldsymbol{\theta},D) \, p(\boldsymbol{\theta}|D) \,,
\end{align}
where $\bm{\theta}$ are the LECs, $p(O_A|\boldsymbol{\theta},D)$ is the probability of predicting $O_A$ given a specific set of LEC values and calibration data $D$, and $p(\boldsymbol{\theta}|D)$ is the LEC posterior originating from our aforementioned Bayesian fit. In this work,
we make the assumption that the EFT truncation uncertainties enter the predictions for $O_A$ only through their effect on the inferred LEC posterior, i.e., $p(O_A|\boldsymbol{\theta},D) = \delta[O_A-O_A^{\mathrm{AFDMC}}(\boldsymbol{\theta})]$. A complete analysis would involve building the full hierarchical error model where correlations between the EFT truncation errors for $O_A$ and $D$ are incorporated. Clearly, this is an outstanding problem, although see Ref.~\cite{Jiang_Forssen_Djarv_etal_2024, Jiang_Forssen_Djarv_etal_2024a} for an example of another approach. We plan to address more sophisticated error models in future work.

\section{Emulators for light nuclei}\label{Xsec3-3}
Evaluating the likelihood function in Eq.~(\ref{likelihood}) by directly solving the Faddeev equations and calculating the $^4$He $r_\text{ch}$ with AFDMC for each LEC sample is computationally prohibitively expensive.
To sidestep this, we turn to machine learning to build emulators, which hold the promise of being able to accurately replicate expensive many-body calculations for a fraction of the computational cost
\cite{Frame_He_Ipsen_etal_2018,Ekstrom_Hagen_2019,Konig_Ekstrom_Hebeler_etal_2020,Melendez_Drischler_Garcia_etal_2021,Wesolowski_Svensson_Ekstrom_etal_2021, Djarv_Ekstrom_Forssen_etal_2022,Bonilla_Giuliani_Godbey_etal_2022,Hu_Jiang_Miyagi_etal_2022,Giuliani_Godbey_Bonilla_etal_2023,Drischler_Melendez_Furnstahl_etal_2023,Odell_Giuliani_Beyer_etal_2024,CompanysFranzke_Tichai_Hebeler_etal_2024,Belley_Yao_Bally_etal_2024,Jiang_Forssen_Djarv_etal_2024a,Duguet_Ekstrom_Furnstahl_etal_2024, Somasundaram_Svensson_De_etal_2024,Somasundaram_Armstrong_Giuliani_etal_2025,Armstrong_Giuliani_Godbey_etal_2025,Cheng_Godbey_Niu_etal_2025,Curry_Kozar_Gezerlis_2025,CompanysFranzke_Tichai_Hebeler_etal_2025}.

Here, we employ two kinds of emulators.
Eigenvector continuation \cite{Frame_He_Ipsen_etal_2018,Duguet_Ekstrom_Furnstahl_etal_2024} projects a Hamiltonian into a subspace spanned by training vectors $|\psi_j\rangle$, which are given by wave functions for several LEC sets.
In our case, $|\psi_j\rangle$ are the Faddeev wave functions.
The subspace-projected Hamiltonian and norm matrices, $\tilde{H} = \langle \psi_i | H | \psi_j \rangle$ and $\tilde{N} = \langle \psi_i | \psi_j \rangle$ are then used in the generalized eigenvalue problem, $\tilde{H} \bm{v} = \tilde{E} \tilde{N} \bm{v}$, where $\bm{v}$ and $\tilde{E}$ are the EC eigenvectors and energies. The lowest eigenstate is the variational estimate for the system's ground-state energy.
In practice, one directly computes the elements of the overlap matrices $\tilde{H}$ and $\tilde{N}$ for all training points, and diagonalizes the $N_{\text{train}} \times N_{\text{train}}$ matrix to find the EC prediction for a given set of LECs.
One of the drawbacks of using EC emulators is the required knowledge of the many-body wave function.
This makes EC particularly unsuitable for use with quantum Monte Carlo methods, but see Ref.~\cite{Sarkar_Lee_Meissner_2023,Somasundaram_Armstrong_Giuliani_etal_2025} for possible solutions. To ensure our EC matrices are not ill-conditioned, we choose additional training points from a set of computed points such that the eigenvalue condition number \cite{Gezerlis_2023} is not appreciably increased.

\begin{figure}[!t]

\centerline{\includegraphics[width=\columnwidth]{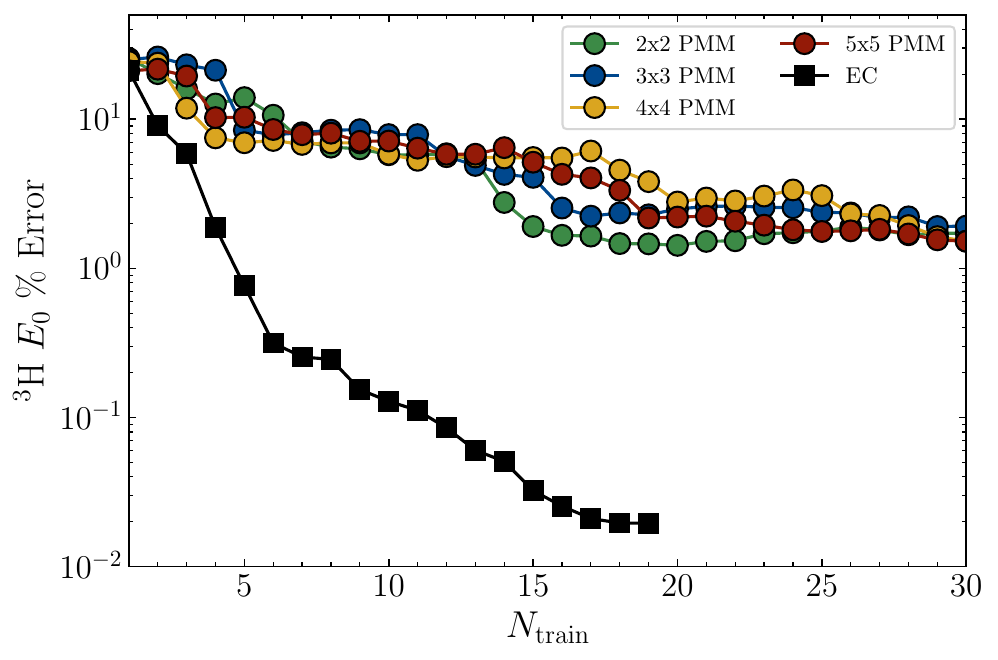}}
\caption{Percent error of the emulators for the $^3\text{H}$ energy as a function of the number of training points.
    This is computed by comparing the emulator predictions against a set of 75 exact Faddeev calculations, which remain fixed for all $N_{\text{train}}$ and PMM dimension.
    Both the PMM and EC emulators are tuned to reproduce the $^3\text{H}$ energy across the 11-dimensional LEC space at N$^2$LO. The training points are chosen uniformly from the LEC prior distributions. For the EC it was also necessary to choose training points from this set such that the matrices did not become ill-conditioned.}
\label{fig:ntrain}
\end{figure}

\begin{figure}[b!]
\centerline{\includegraphics[width=\columnwidth]{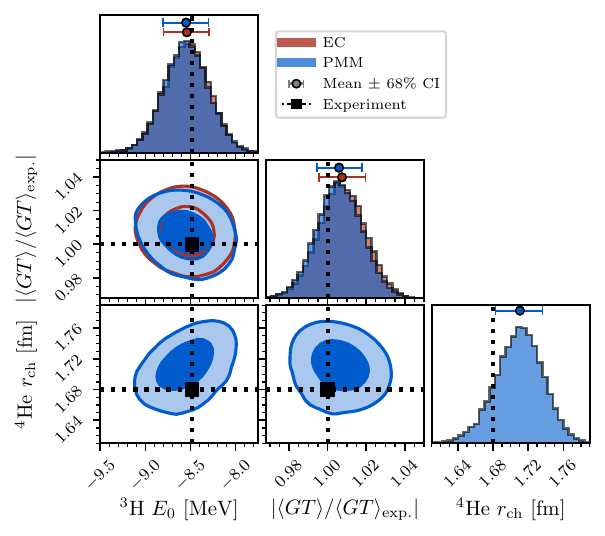}}
\caption{Posteriors for the $^3$H energy, GT matrix element ratio to experiment, and $^4$He charge radius used to constrain the 3N couplings $c_D$ and $c_E$ in our final Bayesian fit (fit 2).
    For the $^3$H properties, we compare predictions for both the EC and PMM emulators, and find that despite the EC's significantly smaller emulator error (see Fig.~\ref{fig:ntrain}), differences between the resulting distributions are negligible. The shaded regions in the correlation plots cover 50\% and 90\% of the total probability. Experimental values used for the fits are from \citet{Wang_Huang_Kondev_etal_2021, Gazit_Quaglioni_Navratil_2009, Krauth_Schuhmann_Ahmed_etal_2021}.}
    \label{fig:fit-observables}
\end{figure}

\begin{figure*}
\centerline{\includegraphics[width=\textwidth]{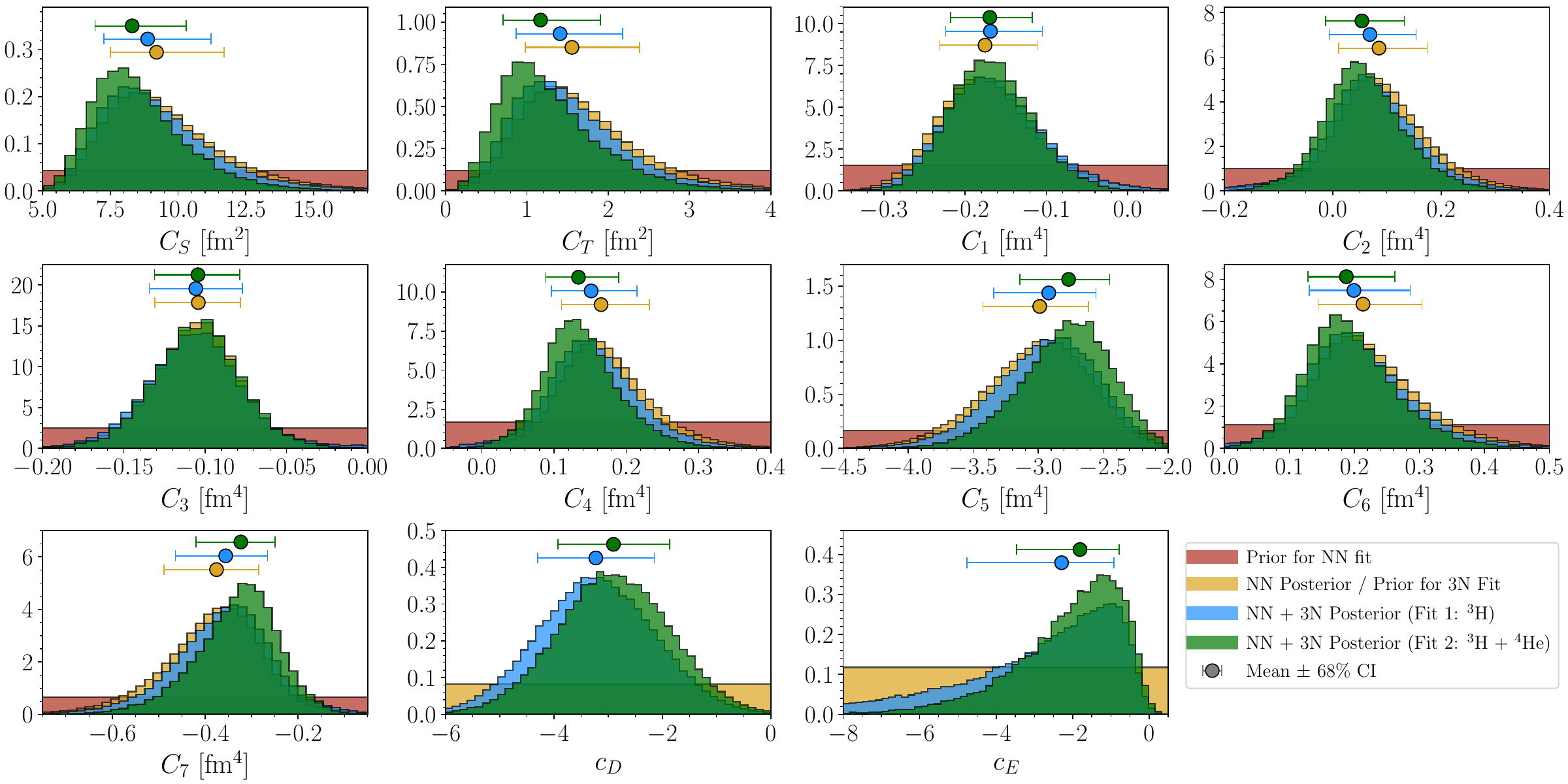}}
\caption{Normalized posterior distributions for all LECs at N$^2$LO with $R_0=0.6\ \text{fm}$ and the $\text{E}_{\tau}$ form of the 3N interaction~\cite{Lynn_Tews_Carlson_etal_2016}.
    The initial uniform prior for the NN LECs is shown in red.
    The posterior of the NN Bayesian fit from Ref. \cite{Somasundaram_Lynn_Huth_etal_2024} together with uniform priors for $c_D$ and $c_E$ is shown in yellow, and the new result for the full posterior for all 11 LECs is shown in blue (fit 1 to only $^3$H $E_0$ and GT matrix element) and green (fit 2 with the additional $^4$He charge radius constraint).}
    \label{fig:full-posterior}
\end{figure*}

To allow for the inclusion of the $^4$He charge radius into our fit, and to eventually propagate the theoretical uncertainties to AFDMC predictions of light nuclei, we turn to an alternative approach.
The PMM is a recently developed machine-learning algorithm \cite{Cook_Jammooa_Hjorth-Jensen_etal_2025} that has proven very useful for emulating many-body nuclear physics calculations \cite{Cook_Jammooa_Hjorth-Jensen_etal_2025,Somasundaram_Armstrong_Giuliani_etal_2025, Reed_Somasundaram_De_etal_2024, Somasundaram_Svensson_De_etal_2024, Armstrong_Giuliani_Godbey_etal_2025,Yu_Miyagi_2025, Curry_Kozar_Gezerlis_2025}.
Similarly to EC, the basic premise is to replace the high-fidelity calculation with an approximate model defined in a subspace.
The model is given by the matrix
\begin{align}
	A(\bm{c}) = A_0 + \textstyle\sum_i c_i A_i\,,
\end{align}
where $A_0$ is a diagonal matrix, the $A_i$ are symmetric matrices, and the $c_i$ are the control parameters (i.e., LECs) of the system.
The dimensionality of the $A$ matrices is a hyperparameter which controls the size of the subspace at the cost of an increased number of fitting parameters.
This is similar to EC, where the subspace dimension is set by the number of training points.
In our case, we represent the Hamiltonian as
\begin{align} \label{eq:PMM-H2}
	\hat{H} &= H_0 + C_S H_S + C_T H_T + \sum_{i=1}^7 C_i H_i + c_D H_D + c_E H_E,
\end{align}
where $C_S$ and $C_T$ are the leading-order NN LECs, the $C_i$'s are the 7 NN LECs that enter at next-to-leading order, and $c_D$ and $c_E$ are the N$^2$LO 3N LECs.
$H_0$ captures the kinetic energy operator and all pion-exchange contributions whose LECs are kept fixed.
The elements in the eleven $H$ matrices can then be fit so that the lowest eigenvalue of $\hat{H}$ reproduces the observable of interest for $N_{\rm train}$ training sets of the LECs.

To do this, we follow Ref.~\cite{Armstrong_Giuliani_Godbey_etal_2025} and first tune the matrix elements of Eq.~(\ref{eq:PMM-H2}) to reproduce the Faddeev or AFDMC results. We initially start with a single training point (i.e., one given set of couplings) and minimize the difference between the high-fidelity calculation and emulator output using a least-squares minimization procedure. As seen in Fig.~\ref{fig:ntrain}, this typically leads to a very poor description of our validation set (high-fidelity calculations that are not used to train the emulator). However, we then add a second training point (an additional set of LECs), and retune the PMM matrix elements starting from their previous values so that both exact results are reproduced.
This process continues, successfully introducing additional training points and re-tuning our matrix elements, until either the desired precision has been reached or adding more training points only gives negligible improvement, see Fig.~\ref{fig:ntrain}. The sets of LECs (training points) are chosen uniformly from the respective prior distributions.
This highlights a considerable strength of the PMM approach, in that additional training points are easy to add without the need for additional overlap computations as in the case of EC.
To improve the accuracy of the PMM when extrapolating, we also ensure that the training set includes the smallest and largest values of the high-fidelity results \cite{Armstrong_Giuliani_Godbey_etal_2025}.

When varying all 11 LECs we find that our EC emulator is able to reproduce the exact $^3$H energy with an average emulator uncertainty of $\sim 0.02\%$, while the PMM achieves $\sim 1\%$, see Fig.~\ref{fig:ntrain}. The errors are determined by comparing the emulator predictions against a fixed set of test points that are separate from the training points.
Our PMM error is comparable to other recent works applying PMMs to quantum Monte Carlo calculations of neutron matter (with fewer LECs and only NN interactions) \cite{Armstrong_Giuliani_Godbey_etal_2025}.
Similarly, our PMM emulator for the charge radius of $^4$He achieves emulator errors of $\sim 1\%$.
For the beta-decay GT matrix element of $^3$H, the PMM and EC emulators perform similarly well, achieving emulator errors of $\sim 0.2\%$.
This is mainly because the GT matrix element's dominant dependence is on $c_D$, i.e., only one parameter, which aids the performance of the PMM.
While we find that the average emulator error for the triton energy is two orders of magnitude smaller for EC than for the PMM, we find only negligible differences in the final propagated posterior distributions, see Fig.~\ref{fig:fit-observables}.
Hence, regardless of the emulator used in the Bayesian likelihood calculation, we also find negligible differences in the posterior distributions of the 11 LECs that are shown in Fig.~\ref{fig:full-posterior}.

In Fig.~\ref{fig:full-posterior}, we show, at the level of the LECs, the result of two different fits carried out in this work.
Fit 1 only considers the triton energy and GT matrix element to constrain $c_D$ and $c_E$, while fit 2 also includes the $^4$He charge radius.
For both fits, the final posterior for the NN LECs is heavily dominated by NN scattering phase shifts and does not change considerably when $A=3,4$ observables are included. This differs from the results in Ref.~\cite{Carlsson_Ekstrom_Forssen_etal_2016} which found that for non-local chiral EFT interactions the posterior distributions of the LECs were strongly dependent on whether the NN and 3N terms were fit simultaneously or separately. We attribute this difference to the inclusion of EFT truncation uncertainties in our fits.
We find that by including the $^4$He charge radius in the fit, all 11 LECs are more constrained, and in particular the long tail of the $c_E$ distribution is reduced.
Additionally, fit 2 leads to more reasonable many-body predictions in the helium isotopes.

\section{Uncertainty propagated auxiliary-field diffusion Monte Carlo}\label{Xsec4-4}
With the full posterior at N$^2$LO in hand, we are able to fully propagate the uncertainties from our local chiral EFT interactions to many-body calculations of the helium isotopes using AFDMC~\cite{Schmidt_Fantoni_1999}.
AFDMC is a powerful computational method that solves the many-body Schr\"{o}dinger equation by projecting out the ground-state wave function of a system through an imaginary-time propagator acting on a trial wave function, $|\psi_0\rangle = \lim_{\tau \rightarrow \infty} e^{-H\tau} |\Psi_T\rangle$.
The trial wave function used in our calculations is of a shell-model-like form and includes two- and three-body correlation functions multiplying a sum of Slater determinants \cite{Lonardoni_Gandolfi_Lynn_etal_2018}.

To build PMM emulators for AFDMC, we calculate 30 training points for each $^3\text{He}$, $^4\text{He}$, and $^6\text{He}$.
This involves considerable computational cost as each set of LECs requires an independent careful optimization of the wave function and transient estimates are required to filter out remaining contaminations from the trial wave function.
Details of the wave function optimization and transient estimate can be found in Ref.~\cite{Lonardoni_Gandolfi_Lynn_etal_2018}.
The total computational cost of generating the AFDMC training data was approximately 5 million CPU-hours.

Once the PMMs are trained, we are able to compute AFDMC energies for the entire posterior distribution, evaluating all observables for $\sim 2.5 \times 10^5$ samples in a matter of minutes on commodity hardware.
In Fig.~\ref{fig:He}, we show our first AFDMC predictions for light nuclei with all theoretical and experimental uncertainties propagated from the interaction to the observables.
Within the 68\% credible interval, our results are in very good agreement with experimental values for the energies and with previous AFDMC calculations of these systems~\cite{Lonardoni_Gandolfi_Lynn_etal_2018}. It is worth noting that even though the interaction used in Ref.~\cite{Lonardoni_Gandolfi_Lynn_etal_2018} employs a different regulator than the one used in our current calculations, they both correspond to reasonably large momentum cutoffs. The $R_0=1.0$ fm used in Ref.~\cite{Lonardoni_Gandolfi_Lynn_etal_2018} corresponds to $\Lambda_C \approx 500$ MeV, compared with our calculations which employ a local regulator from Ref.~\cite{Tews_Huth_Schwenk_2018} with $R_0=0.6$ fm, which corresponds to $\Lambda_C = 2/R_0 \approx 660$ MeV.

We find that our predictions for the energies of all three helium isotopes are closer to the experimental values and have tighter credible intervals for fit 2, even though these observables were not included in our fits.
While the central values for our distributions do indicate an unstable $^6$He, our propagated uncertainties result in credible intervals that overlap with a stable $^6$He nucleus.
As we propagate the Hamiltonian uncertainty directly, we are also able to extract the correlations between observables for different helium isotopes.
As expected, we find a reasonably strong correlation between the energies of $^3$He and $^4$He.
In contrast, $^6$He exhibits no correlation with these lighter isotopes.
We expect this to be due to the halo structure of $^6$He \cite{Navratil:2026hbo}.
In the future, we plan to pursue additional studies of halo nuclei using AFDMC calculations with propagated\break uncertainties.

\section{Summary}\label{Xsec5-5}
We have implemented emulators for the solutions to the Faddeev equations for both the energy and the Gamow-Teller matrix element of $^3$H, and for AFDMC calculations of the $^4$He charge radius, to carry out a Bayesian determination of the full posterior distribution for all two- and three-body low-energy couplings at N$^2$LO.
We find that the NN LECs in our Hamiltonian are strongly dominated by NN scattering.
We implemented EC and PMM emulators and found their performance to be comparable.
The PMM is needed when explicit information regarding the wave function is missing, which is the case for our AFDMC calculations.
We then used PMM emulators trained on AFDMC calculations of helium isotopes to propagate experimental and theoretical uncertainties inherent in our interactions to the many-body observables.
As a result, we are able to provide robust uncertainty estimates for our AFDMC predictions of the energies, as well as the correlations between different nuclei along the isotopic chain.
This work serves as a framework for future \textit{ab initio} studies of nuclei where all sources of uncertainty are treated and propagated appropriately.

\section*{Acknowledgments}
The authors would like to thank C.L. Armstrong and D. Lonardoni for insightful discussions. 
The work of R.C. and A.G. was supported by the Natural Sciences and Engineering Research Council (NSERC) of Canada and the Canada Foundation for Innovation (CFI).
The work of K.H. and A.S. was supported in part by the European Research Council (ERC) under the European Union’s Horizon 2020 research and innovation programme (Grant Agreement No.~101020842).
S.G., R. S., and I.T. were supported by the U.S. Department of Energy through Los Alamos National Laboratory (LANL). 
LANL is operated by Triad National Security, LLC, for the National Nuclear Security Administration of U.S. Department of Energy (Contract No.~89233218CNA000001).
S.G., R. S., and I.T.  were also supported by the U.S. Department of Energy, Office of Science, Office of Advanced Scientific Computing Research, Scientific Discovery through Advanced Computing (SciDAC) NUCLEI program.
R.C., S.G., R.S., and I.T. were also supported by the Laboratory Directed Research and Development (LDRD) program of LANL under project number and 20230315ER.
R.S. and I.T. are also supported by the LDRD program of LANL under project number 20260260ER.

Computational resources have been provided by Compute Ontario through the Digital Research Alliance of Canada, the Los Alamos National Laboratory Institutional Computing Program, which is supported by the U.S. Department of Energy National Nuclear Security Administration under Contract No. 89233218CNA000001, and by the National Energy Research Scientific Computing Center (NERSC), which is supported by the U.S. Department of Energy, Office of Science, under Contract No. DE-AC02-05CH11231.

\bibliography{bib}

\end{document}